# Enhanced High-Harmonic Generation from an All-Dielectric Metasurface

**Hanzhe Liu[1,2], Cheng Guo[3,4], Giulio Vampa[1], Jingyuan Linda Zhang[3,4], Tomas Sarmiento[4,5], Meng Xiao[4], Philip H. Bucksbaum[1,2,3,6], Jelena Vučković[3,4,5], Shanhui Fan[3,4,5] and David A. Reis[1,3,6]**

*[1]Stanford PULSE Institute, SLAC National Accelerator Laboratory, Menlo Park, California, 94025, USA*

*[2]Department of Physics, Stanford University, Stanford, California, 94305, USA*

*[3]Department of Applied Physics, Stanford University, Stanford, California, 94305, USA*

*[4]E. L. Ginzton Laboratory, Stanford University, Stanford, California 94305, USA*

*[5]Department of Electrical Engineering, Stanford University, Stanford, California 94305, USA*

*[6]Department of Photon Science, Stanford University, Stanford, California 94305, USA*

**The recent observation of high-harmonic generation from solids[1–8] creates a new possibility for engineering fundamental strong-field processes by patterning the solid target with subwavelength nanostructures[9,10]. All-dielectric metasurfaces exhibit high damage thresholds and strong enhancement of the driving field[11–20], making them attractive platforms to control high-harmonics and other high-field processes at nanoscales. Here we report enhanced non-perturbative high-harmonic emission from a Si metasurface that possesses a sharp Fano resonance resulting from a classical analogue of electromagnetically induced transparency. Harmonic emission is enhanced by more than two orders of magnitude compared to unpatterned samples. The enhanced high harmonics are highly anisotropic with excitation polarization and are selective to excitation wavelength due to its**

**resonant feature. By combining nanofabrication technology and ultrafast strong-field physics, our work paves the way for designing new compact ultrafast photonic devices that operate under high intensities and short wavelengths[21].**

High-harmonic generation (HHG) was first reported in rare gas atoms[22]. It is regarded as one of the fundamental processes in strong-field physics and lies at the heart of attosecond photonics[21]. Recent observations of HHG in solids[1–8] provide a new way to investigate novel strong-field photonic applications that cannot be realized in gases. For example, due to the high density of solids, high-harmonics can be generated and controlled by patterning the solid target with sub-wavelength nanostructures in a desired fashion[9,10,23]. The effect of nanostructures on HHG is two-fold: first, each individual nanoscale feature interacts with and scatters fundamental light in a non-trivial way depending on its geometry; and, second, HHG emission profiles in the far field can be controlled by arranging the location of individual nanostructures. For example, harmonics have been produced in Fresnel zone plates producing a diffraction-limited spot in the far-field[9].

One of the challenges in realizing practical devices for ultrafast strong-field nanophotonics is that the nonlinear response for most solids is intrinsically weak, especially when the excitation photon energy is much smaller than the bandgap, which is typically the case in solid-HHG experiments[1–8]. Phase matching in this context is difficult for above-gap harmonics as the attenuation length for high harmonics is normally much shorter than the coherence length due to strong absorption during propagation in bulk[1]. A promising solution is to design field-enhancement nanostructures on a surface[10,14,16–18,23,24]. Plasmon-enhanced HHG has been reported from metallic antennas deposited on Si[10] and sapphire[23]. However, the damage threshold of plasmonic-based metal structures is extremely sensitive to the fabrication quality and the high degree of confinement in the vicinity of sharp features such as edges or corners, significantly limits the overall signal enhancement[10,24,25].

In this letter, we report enhanced high harmonics from an all-dielectric Si metasurface on a sapphire substrate comprising optically resonant dielectric nanostructures. We demonstrate that when the metasurface is resonantly excited, the high-harmonic signal is orders of magnitude higher compared to an unpatterened Si film at moderate driving intensities. Figure 1**a** shows an SEM image of the metasurface consisting of dipolar antennas (bars) and disk resonators fabricated from a 225 nm thick single crystal Si film grown over a 0.53 mm thick sapphire substrate. This type of structure is known to host a Fano resonance resulting from a classical analogue of electromagnetically induced transparency (EIT) when the excitation polarization is along the bar[14,15]. Figure 1**b** displays the photonic level scheme for the Si bar antenna and disk resonator. Upon normal incident illumination, an electric dipolar mode in the bar is directly excited and is illustrated as the transition between the ground state |0> and the excited states |1> of the bar. While the excitation field inside the bar decays into free space rapidly, part of the mode couples to a magnetic dipole mode supported by the disk that does not radiate directly. The coupling between the dipolar bar mode and the 'dark' disk mode allows the disk to serve as a reservoir to confine the fundamental electric field after the excitation, and to transfer energy back to the dipolar mode at a much later time compared to the excitation pulse duration. This energy transfer process due to mode coupling between bars and disks is illustrated as transition (|0> → |1> → |2> → |1>) in Fig. 1**b**. The interference between the two excitation pathways, namely the direct excitation of the dipolar mode inside the bar (|0> → |1>) and the indirect excitation due to the coupling between bars and disks (|0> → |1> → |2> → |1>), represents a Fano interference process and shows in a sharp EIT peak in the transmission spectrum [14,15,26]. Figure 1**c** displays

the simulated and measured transmission spectrum for our device. An EIT peak is observed in experiment around 2320 nm and agrees well with simulation.

One important feature for EIT is that it gives rise to greatly enhanced nonlinear light-matter interaction in the spectral region of the resonance[27]. Enhanced frequency conversion from EIT has been observed in atomic systems[27] and very recently, in photonic devices in the perturbative regime for third harmonic generation[14]. Here we excite our metasurface with infrared radiation tuned to the EIT resonance and report greatly enhanced solid HHG in the non-perturbative regime, where the driving field inside the material is close to the intrinsic field between atomic sites in solids.

We generate solid-state high harmonics using linearly polarized, 1kHz repetition rate, 70 fs pulses centered at 2320 nm, which are well below the indirect band gap of silicon. The laser pulses are focused on the device down to a waist diameter of ~160 μm, covering an area consisting of ~15600 unit cells of the metasurface. The periodic pattern diffracts the high harmonics. The zeroth order diffracted harmonics are collected and measured in a UV-VIS spectrometer, and are compared to those emitted from unpatterned regions on the same film under identical illumination and detection conditions.

Figure 2 shows the high-harmonic spectrum produced by the metasurface at an incident excitation intensity of 0.071 TW/cm$^2$. When the fundamental field is polarized along the bar (red line), odd harmonics from 5$^{th}$ to 11$^{th}$ are observed, while no harmonics above 5$^{th}$ are observed in unpatterned Si (grey line). At this intensity, the 5$^{th}$ harmonic is enhanced by a factor of 30 compared to the unpatterned film, even though only the zero diffraction order is collected and the total area of structured Si is only 47% of the unpatterned film under same illumination spot. When the polarization of the excitation pulse is perpendicular to the bar (green line), a configuration where EIT resonance is not supported, only 5$^{th}$ and 7$^{th}$ harmonics are observed. In this case, the yield for the 5$^{th}$ harmonic is similar to the unpatterned film and a 7$^{th}$ harmonic peak is seen above the noise level but reduced by a factor of 12 compared to the same harmonic yield in parallel configuration.

To further determine the HHG enhancement for individual harmonics, we measured the high-harmonic yield as a function of the incident excitation intensity $I$ (Fig. 3). The polarization of the excitation pulse is parallel to the bars. As illustrated in Fig. 3, all three observable harmonics, 5$^{th}$, 7$^{th}$ and 9$^{th}$, show large enhancements in the spectrally integrated yield compared to the unpatterned Si film, especially at relatively low excitation intensities. The enhanced high-harmonic emission indicates that the field energy density concentrated in the nanostructure is higher than that in the unpatterned area. This claim is further supported by the different scaling behavior of the metasurface-enhanced HHG and HHG from bare Si film. The high-harmonics yield from the metasurface as a function of excitation intensity deviates from the perturbative scaling and exhibits a gradual saturation. Note that 5$^{th}$ harmonic from bare film is still perturbative ($I^5$) under similar excitation.

After saturation is reached by slowly increasing the driving intensity, the high-harmonic yield does not return to its former value when the intensity is lowered. This suggests progressive damage of the structures and further indicates that the field is significantly enhanced by the device. By comparing high-harmonic yield between different orders, we further note that higher order harmonics are much more enhanced than the low order ones. For instance, the enhancement for 7$^{th}$ harmonic at around 0.1 TW/cm$^2$ is more than two orders of magnitude, while the 5$^{th}$ harmonic under same intensity is enhanced by about one order of magnitude.

To further reveal the resonant nature of EIT enhanced high-harmonics, we measured the high-harmonic spectrum under various excitation wavelengths. Figure 4**a** displays the spectrum of the 5$^{th}$ harmonic generated by three different excitation wavelengths at 2330 nm, 2280 nm and 2209 nm respectively under the same intensity of 0.07 TW/cm$^2$. While 2330 nm overlaps well with the EIT resonance, 2280 nm and 2209 nm excitation are blue-shifted from the EIT peak. The effects of detuning on HHG are two-fold. First, since the excitation bandwidth is much broader than the EIT line-width, when the detuning is slight, part of the excitation pulse spectrum is still resonantly coupled to the metasurface. As a result, with small detuning, there is an enhanced HHG peak that is locked to the 5$^{th}$ harmonic of the EIT resonance frequency rather than the excitation laser frequency, denoted as the red arrow in Fig. 4**a**. The intensity of this peak decreases monotonically with detuning of the excitation pulse. Second, as the excitation wavelength is detuned away from resonance, a second HHG peak outside the EIT resonance emerges, denoted as green and blue arrows in Fig. 4**a**. This corresponds to the 5$^{th}$ harmonic of the excitation beam at its center wavelength. This peak shifts its center position as the excitation wavelength is detuned and is not enhanced since the center wavelength of the excitation pulse and the EIT resonance are no longer overlapped under detuning. Figure 4**b** shows that the spectrally integrated HHG yield decreases as the excitation laser is detuned from resonance.

Our demonstration of HHG on an all-dielectric metasurface has several important implications. First, compared to plasmonic nanostructures, an all-dielectric metasurface has the advantage of high transparency for the fundamental pulse and higher damage threshold as a high harmonic source, which is crucial for XUV HHG process in solids, where the excitation intensity is normally one order of magnitude higher than the excitation intensity reported in current work[2,6,8]. One way to push the high-energy cutoff into the XUV regime might be to replace the Si metasurface with a larger bandgap material, known to generate VUV/XUV harmonics (e.g. ZnO[1], MgO[8]), and to be robust under higher excitation intensity.

Second, unlike most plasmonic-based nanostructures, dielectric photonic structures can exhibit sharp resonances[11–20] that are narrower than the bandwidth of typical excitation pulses used in solid HHG experiments[1–8]. This opens a novel avenue to control the HHG process in both temporal and frequency domain. A sharp resonance in the frequency domain is equivalent to a long lifetime of confinement of light in a high Q resonator, which can lead to a delayed emission of high-harmonics in the time domain. This delay can be tuned by adjusting the Q of the metasurface, which can be achieved simply by changing the distance between the disks and bars in the current design[15]. Furthermore, when the metasurface is excited with a broadband transform-limited input pulse, a destructive interference is induced between spectral components that are blueshifted and redshifted from the resonant wavelength, limiting the peak transient energy coupled into the metasurface[28]. By coherent control of the input pulse, especially by tuning the spectral phase via pulse shaping techniques, one can control the peak field coupled into the dielectric nanostructure and in this case, further increase the extreme nonlinear light-matter interaction strength.

Third, the polarization state of light can be controlled by rearranging the position of individual nanoresonators in an all-dielectric metasurface[13,19,29,30]. Specifically, a high-harmonic beam with circular polarization can be achieved by applying circularly polarized light to pump a metasurface with similar structures but being rotated by 90 degrees every few unit cells. In general, control of both amplitude and phase of the infrared excitation light is possible with anisotropic nanostructures that are rotated periodically[13,19,29,30]. This degree of control is transferred to the high harmonics, thereby enabling spatial manipulation of high-harmonic emission to generate arbitrary polarization states as well as non-zero orbital angular momentum.

In conclusion, we have demonstrated solid-state HHG from an all-dielectric Si metasurface. The high harmonics yield is greatly enhanced by the sharp EIT resonance of the metasurface compared to unpatterened Si film and exhibits a strong anisotropic response. The generation of high-harmonics from an all-dielectric metasurface provides a different approach to manipulate and control the HHG process at nanoscales, which will enable spatio-temporal control of the properties of the high-harmonic beam. By combining nanophotonics and high-field laser physics, fundamental strong field physics processes can be engineered in novel ways and new devices and applications operating under high intensities and short wavelengths can be realized.

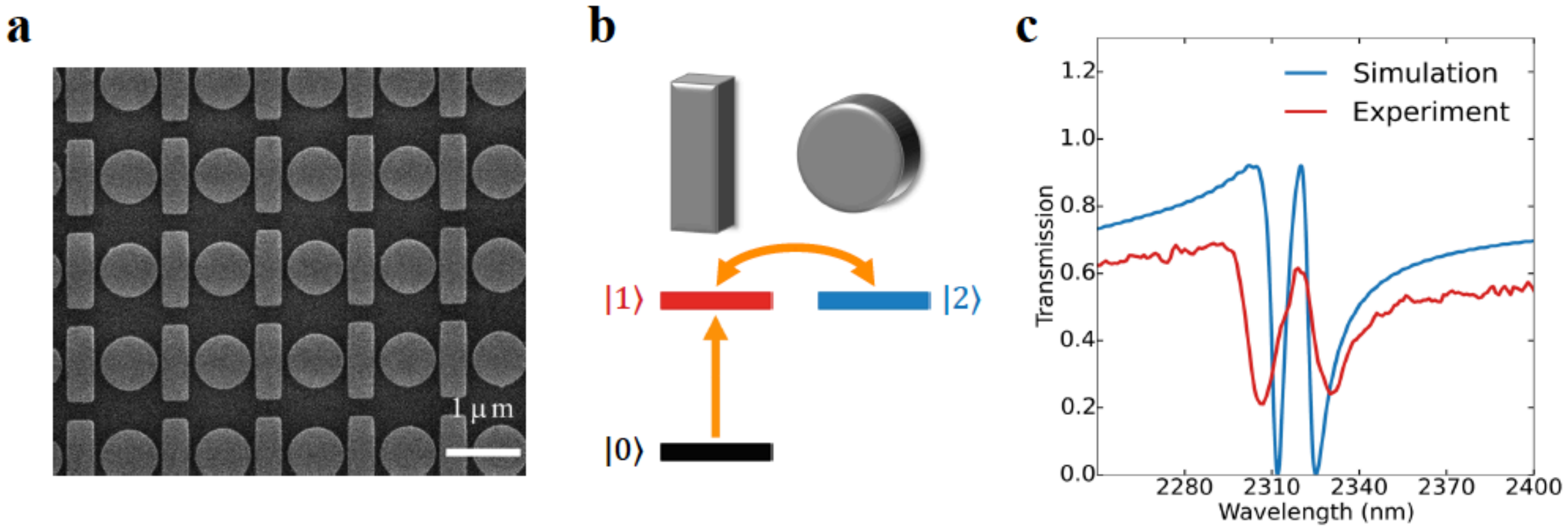

**Figure 1 | Working principle of designed device and resonance characterization. a,** An SEM image of fabricated Si metasurface on sapphire substrate. A single unit cell consists of a dipolar bar antenna and a disk resonator to its right. Geometric parameters of the metasurface are: Si thickness = 225 nm (on top

of sapphire); unit cell period = 1280 nm; length of the bar = 1022 nm; width of the bar = 347 nm; disk radius = 369 nm; distance from bar center to the right disk center = 652 nm. **b,** Level scheme for the mode coupling in a schematic three-level system. States |1> and |2> correspond to the directly excited dipolar mode in the bar and indirectly excited magnetic dipole mode in the disk. EIT resonance results from interference between two pathways (|0> → |1> and |0> → |1> → |2> → |1>). **c,** Experimentally measured (red line) and simulated (blue line) transmission spectrum of the metasurface.

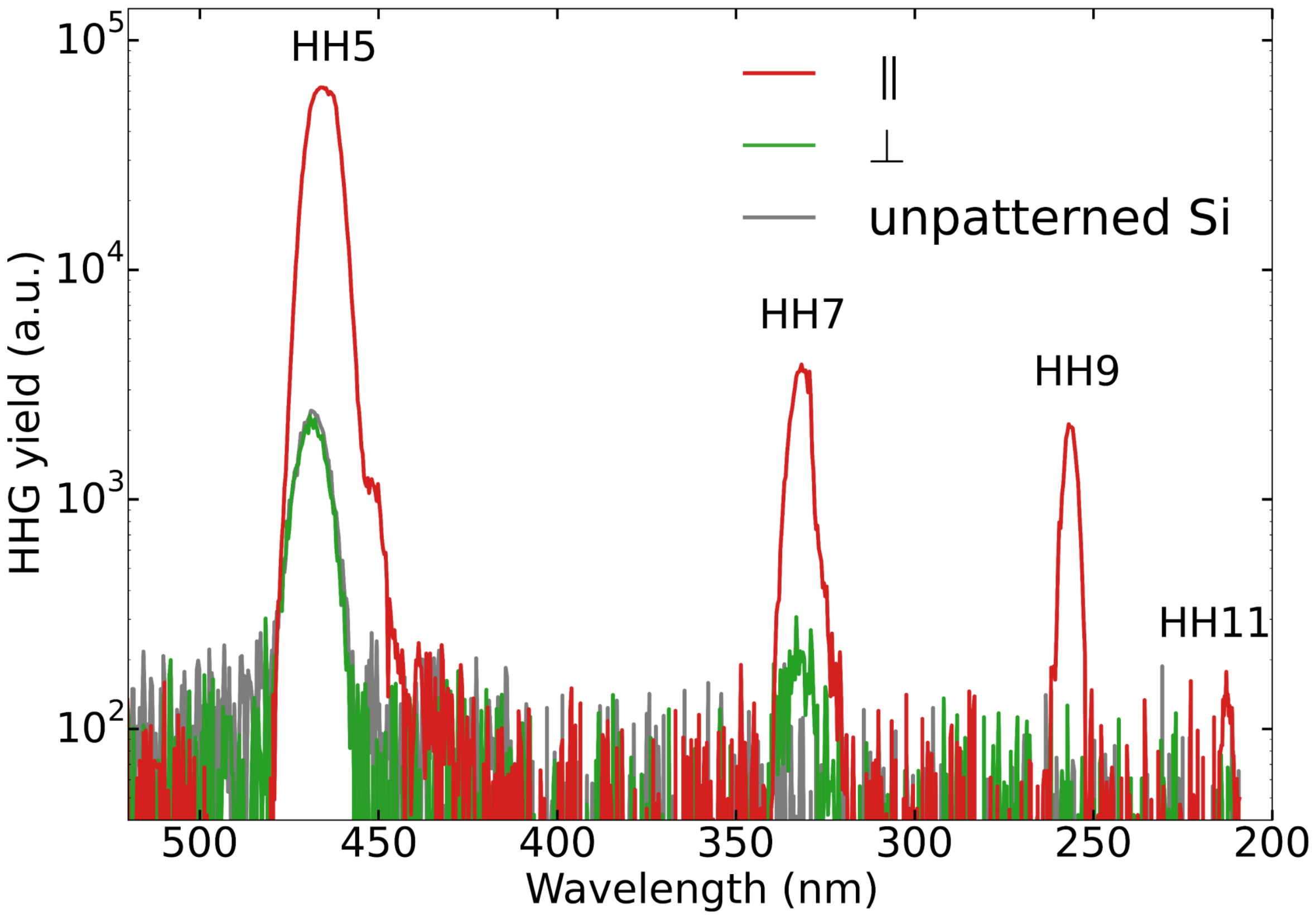

**Figure 2 | High-harmonic spectra from Si metasurface.** High-harmonic spectra from 5$^{th}$ to 11$^{th}$ order when the excitation pulses centered at 2320 nm are polarized parallel (red line) and perpendicular (green line) to the bar on the metasurface. The gray line is the high-harmonic spectrum from unpatterned Si film. The spectra are taken at a vacuum excitation intensity of 0.071 TW/cm$^2$. The reported spectral range is limited by our detection scheme, which cuts off at the short-wavelength edge of the 11$^{th}$ harmonic.

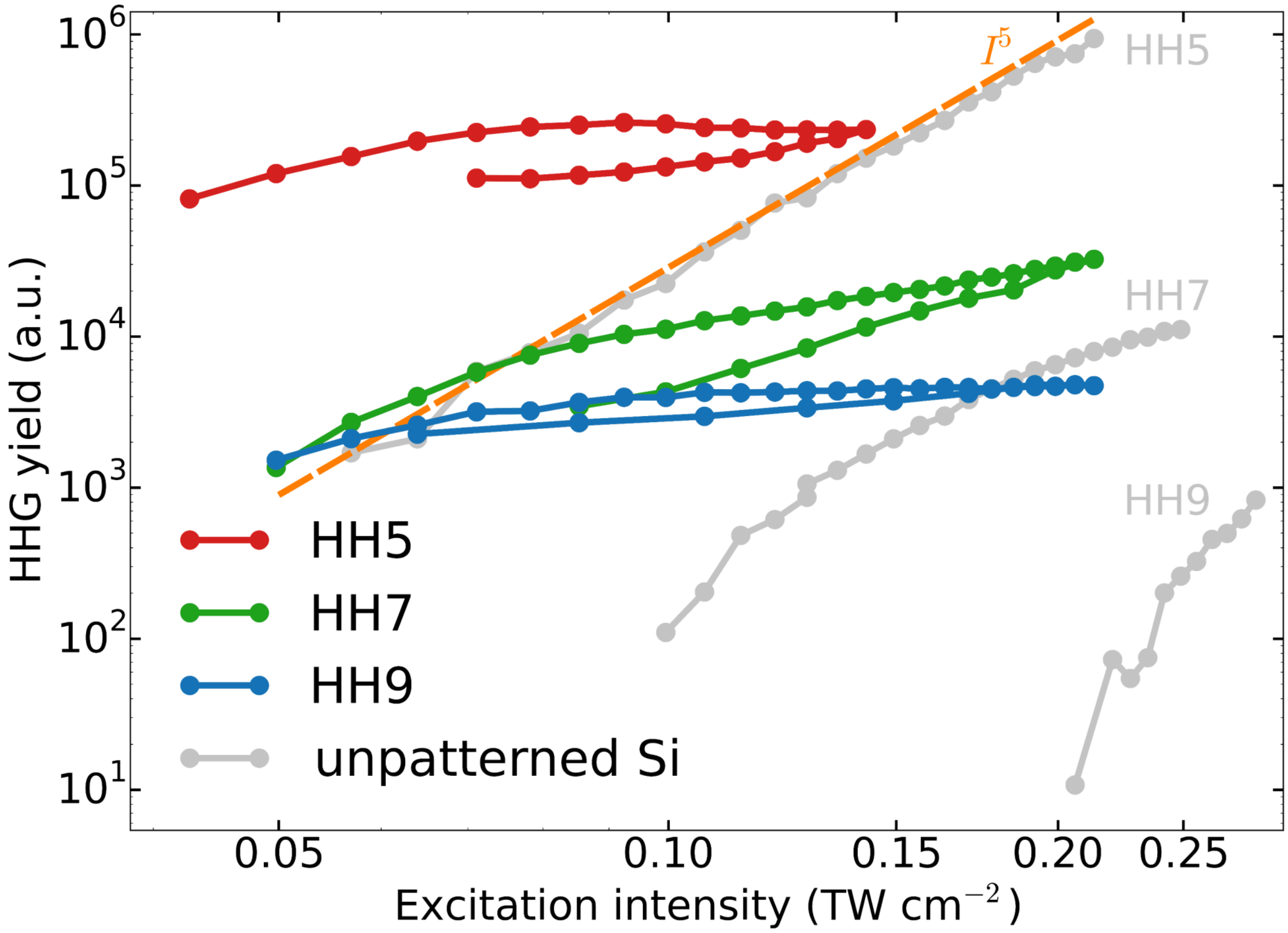

**Figure 3 | Dependence of non-perturbative high-harmonics yield on excitation intensity.** The measured harmonic yield as a function of vacuum peak excitation intensity $I$ for $5^{th}$ (red), $7^{th}$ (green) and $9^{th}$ (blue) harmonic when the excitation pulse centered at 2320 nm is polarized along the bar. HHG yield is measured as the excitation intensity increases (first) and then decreases when the yield slowly saturates. All harmonics from metasurface scale non-perturbatively. The gray circles show the HHG emission from unpatterned Si film for the same three harmonics. The dashed orange line, which scales as $I^5$, is plotted for guiding the eye.

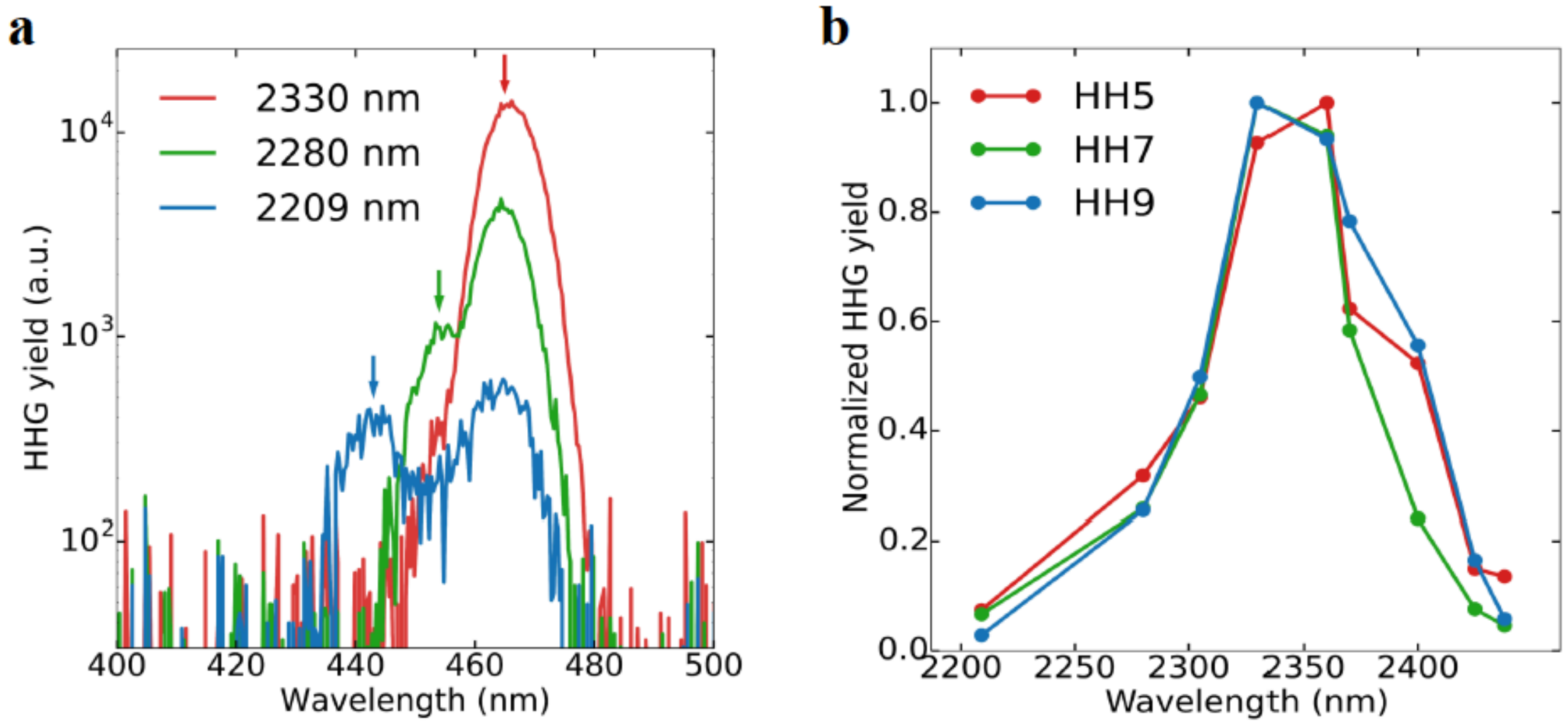

**Figure 4 | Dependence of high-harmonic spectrum on excitation wavelength. a,** $5^{th}$ harmonic spectrum under excitation at 2330 nm (red line), 2280 nm (green line) and 2209 nm (blue line). 2330 nm is resonant with the EIT peak. The red arrow denotes the wavelength of the $5^{th}$ harmonic of the EIT resonance. The green and blue arrows denote the side peaks, which correspond to the $5^{th}$ harmonic for 2280 nm and 2209 nm respectively**. b**, Integrated $5^{th}$ (red dots), $7^{th}$ (green dots) and $9^{th}$ (blue dots) harmonic yield as a function of excitation wavelength.

## Acknowledgements

This project was supported primarily by the Air Force Office of Scientific Research under Grant Nos. FA9550-14-1-0108.

## Author Contributions

H.L. and C.G. contributed equally to this work. H.L. conceived the experiment. C.G. and M.X. performed FDTD simulation. H.L. and J.L.Z. fabricated the device. H.L. and G.V. performed the HHG measurement under supervision of D.A.R.. H.L., C.G, T.S., and J.L.Z. characterized the resonance. All authors contributed to the discussion and preparation of the manuscript.

## Competing financial interests

The authors declare no competing financial interests.

## Methods

**Device fabrication.** The devices were fabricated using a Si on sapphire wafer, with a nominal 225 nm thick device layer and 530 μm thick substrate. A JEOL JBX-6300FS electron-beam lithography system was used to pattern a 320 nm thick AR-P 6200.09 electron-beam resist layer spun on Si surface. A plasma etcher was used to transfer the pattern and etch through Si with $BCl_3/Cl_2/O_2$ chemistry. The resist mask was then stripped via oxygen plasma and the device was cleaned in Piranha.

**Measurement of high harmonics.** To generate high-harmonics, the idler beam of a Ti:sapphire laser pumped optical parametric amplifier (OPA), operating at a 1 kHz repetition rate, is used to normally excite the sample. The center wavelength of the idler is tuned around 2.3 μm and the time duration is around 70 fs. HHG is collected and measured in the transmission geometry with a visible-ultraviolet spectrometer after the sample. To reduce the scattering inside the spectrometer, a UV fused silica prism is placed before the spectrometer to pre-disperse the HHG signal. The reported high-harmonic spectral range is limited by the short-wavelength cutoff of the spectrometer.

## Data availability statement

The data that support the plots within this paper and other findings of this study are available from the corresponding author upon reasonable request.

## References

1. Ghimire, S. *et al.* Observation of high-order harmonic generation in a bulk crystal. *Nat. Phys.* **7,** 138–141 (2011).
2. Luu, T. T. *et al.* Extreme ultraviolet high-harmonic spectroscopy of solids. *Nature* **521,** 498–502 (2015).
3. Vampa, G. *et al.* Linking high harmonics from gases and solids. *Nature* **522,** 462–464 (2015).
4. Hohenleutner, M. *et al.* Real-time observation of interfering crystal electrons in high-harmonic generation. *Nature* **523,** 572–575 (2015).
5. Schubert, O. *et al.* Sub-cycle control of terahertz high-harmonic generation by dynamical Bloch oscillations. *Nat. Photonics* **8,** 119–123 (2014).
6. Ndabashimiye, G. *et al.* Solid-state harmonics beyond the atomic limit. *Nature* **534,** 520–523 (2016).
7. Liu, H. *et al.* High-harmonic generation from an atomically thin semiconductor. *Nat. Phys.* **13,** 262–265 (2016).
8. You, Y. S., Reis, D. A. & Ghimire, S. Anisotropic high-harmonic generation in bulk crystals. *Nat. Phys.* **13,** 345–349 (2016).
9. Sivis, M. *et al.* Tailored semiconductors for high-harmonic optoelectronics. *Science (80-. ).* **357,** 303–306 (2017).
10. Vampa, G. *et al.* Plasmon-enhanced high-harmonic generation from silicon. *Nat. Phys.* **13,** 659–662 (2017).
11. Krasnok, A., Tymchenko, M. & Alù, A. Nonlinear metasurfaces: A paradigm shift in nonlinear optics. *Mater. Today* (2017).
12. Kuznetsov, A. I., Miroshnichenko, A. E., Brongersma, M. L., Kivshar, Y. S. & Luk'yanchuk, B. Optically resonant dielectric nanostructures. *Science (80-. ).* **354,** aag2472 (2016).
13. Li, G., Zhang, S. & Zentgraf, T. Nonlinear photonic metasurfaces. *Nat. Rev. Mater.* **2,** 17010 (2017).
14. Yang, Y. *et al.* Nonlinear Fano-Resonant Dielectric Metasurfaces. *Nano Lett.* **15,** 7388–7393 (2015).
15. Yang, Y., Kravchenko, I. I., Briggs, D. P. & Valentine, J. All-dielectric metasurface analogue of electromagnetically induced transparency. *Nat. Commun.* **5,** 5753 (2014).
16. Shcherbakov, M. R. *et al.* Enhanced third-harmonic generation in silicon nanoparticles driven by magnetic response. *Nano Lett.* **14,** 6488–6492 (2014).
17. Liu, S. *et al.* Resonantly Enhanced Second-Harmonic Generation Using III-V Semiconductor All-Dielectric Metasurfaces. *Nano Lett.* **16,** 5426–5432 (2016).
18. Grinblat, G., Li, Y., Nielsen, M. P., Oulton, R. F. & Maier, S. A. Enhanced third harmonic generation in single germanium nanodisks excited at the anapole mode. *Nano Lett.* **16,** 4635–4640 (2016).
19. Jahani, S. & Jacob, Z. All-dielectric metamaterials. *Nat. Nanotechnol.* **11,** 23–36 (2016).
20. Limonov, M. F., Rybin, M. V., Poddubny, A. N. & Kivshar, Y. S. Fano resonances in photonics. *Nat. Photonics* **11,** 543–554 (2017).
21. Vampa, G., Fattahi, H., Vučković, J. & Krausz, F. Nonlinear optics: Attosecond nanophotonics. *Nat. Photonics* **11,** 210–212 (2017).
22. Ferray, M. *et al.* Multiple-harmonic conversion of 1064 nm radiation in rare gases. *J.*

*Phys. B At. Mol. Opt. Phys.* **21,** L31–L35 (1988).
23. Han, S. *et al.* High-harmonic generation by field enhanced femtosecond pulses in metal-sapphire nanostructure. *Nat. Commun.* **7,** 13105 (2016).
24. Stockman, M. I. Nanoplasmonics : The physics behind the applications Nanoplasmonics : The physics behind the applications. *Phys. Today* **64,** 39–44 (2016).
25. Gramotnev, D. K. & Bozhevolnyi, S. I. Plasmonics beyond the diffraction limit. *Nat. Photonics* **4,** 83–91 (2010).
26. Liu, N., Hentschel, M., Weiss, T., Alivisatos, A. P. & Giessen, H. Three-Dimensional Plasmon Rulers. *Science (80-. ).* **332,** 1407–1410 (2011).
27. Fleischhauer, M., Imamoglu, A. & Marangos, J. P. Electromagnetically induced transparency: Optics in coherent media. *Rev. Mod. Phys.* **77,** 633–673 (2005).
28. Sandhu, S., Povinelli, M. L. & Fan, S. Enhancing optical switching with coherent control. *Appl. Phys. Lett.* **96,** 3–5 (2010).
29. Li, G. *et al.* Continuous control of the nonlinearity phase for harmonic generations. *Nat. Mater.* **14,** 607–612 (2015).
30. Yu, N. & Capasso, F. Flat optics with designer metasurfaces. *Nat. Mater.* **13,** 139–150 (2014).